\documentclass[amsmath,aps,prd,showpacs,a4paper,10pt]{revtex4}
\begin{document}

\newcommand{\be}[1]{\begin{equation}\label{#1}}
\newcommand{\ee}{\end{equation}}
\newcommand{\bea}{\begin{eqnarray}}
\newcommand{\eea}{\end{eqnarray}}
\def\disp{\displaystyle}

\newcommand{\cw}{{_{\rm CW}}}  % only used for the index CW standing for Cheng-Weyl

\begin{titlepage}

\begin{flushright}
astro-ph/0607064
\end{flushright}

\title{\Large \bf Cheng-Weyl Vector Field and its Cosmological Application}

\author{Hao Wei}
\email[\,email address:\ ]{haowei@mail.tsinghua.edu.cn}
\affiliation{Department of Physics and Tsinghua Center for
Astrophysics,
Tsinghua University, Beijing 100084, China\\
Institute of Theoretical Physics, Chinese Academy of Sciences,
P.O. Box 2735, Beijing 100080, China}

\author{Rong-Gen Cai}
\affiliation{Institute of Theoretical Physics, Chinese Academy of Sciences,
P.O. Box 2735, Beijing 100080, China}

\begin{abstract}
\vspace{5mm} \centerline{\bf ABSTRACT}\vspace{2mm}
 Weyl's idea on scale invariance was resurrected by Cheng in 1988.
 The requirement of local scale invariance leads to a completely new
 vector field, which we call the ``Cheng-Weyl vector field''. The
 Cheng-Weyl vector field couples only to a scalar field and the
 gravitational field naturally. It does not interact with other
 known matters in the standard model of particle physics. In the
 present work, the (generalized) Cheng-Weyl vector field coupled
 with the scalar field and its cosmological application are
 investigated. A mixture of the scalar field and a so-called
 ``cosmic triad'' of three mutually orthogonal Cheng-Weyl vector
 fields is regarded as the dark energy in the universe. The
 cosmological evolution of this ``mixed'' dark energy model is
 studied. We find that the effective equation-of-state parameter of
 the dark energy can cross the phantom divide $w_{de}=-1$ in some
 cases; the first and second cosmological coincidence problems can
 be alleviated at the same time in this model.
\end{abstract}

\pacs{95.36.+x, 45.30.+s, 98.80.-k}

\maketitle

\end{titlepage}

\renewcommand{\baselinestretch}{1.6}

%\setcounter{page}{1}

%============================= section 1 ===================================

\section{\label{sec1} Introduction}
Since the discovery of present accelerated expansion of our
universe~\cite{r1,r2,r3,r4,r5, r6,r59}, dark energy~\cite{r7} has
been an active field in modern cosmology. One of the puzzles of the
dark energy problem is the (first) cosmological coincidence problem,
namely, {\em why does our universe begin the accelerated expansion
recently? why are we living in an epoch in which the energy
densities of dark energy and dust matter are comparable?} In order
to give a reasonable interpretation to the (first) cosmological
coincidence problem, many dynamical dark energy models have been
proposed, such as quintessence~\cite{r8,r9},
phantom~\cite{r10,r11,r12}, k-essence~\cite{r13,r62} etc.

The equation-of-state parameter (EoS) of dark energy $w_{de}\equiv
p_{de}/\rho_{de}$ plays a central role in observational cosmology,
where $p_{de}$ and $\rho_{de}$ are its pressure and energy density
respectively. Recently, by fitting the observational data, marginal
evidence for $w_{de}(z)<-1$ at redshift $z<0.2\sim 0.3$ has been
found~\cite{r14,r15,r16}. In addition, many best-fits of the present
value of $w_{de}$ are less than $-1$ in various data fittings with
different parameterizations (see~\cite{r17} for a recent review).
The present observational data seem to slightly favor an evolving
dark energy with $w_{de}$ crossing $-1$  from above to below in the
near past~\cite{r15}. Obviously, the EoS of dark energy $w_{de}$
cannot cross the so-called phantom divide $w_{de}=-1$ for
quintessence or phantom alone. Although at first glance, it seems
possible for some variants of k-essence to give a promising solution
to cross the phantom divide, a no-go theorem, shown in~\cite{r18},
shatters this kind of hope. In fact, it is not a trivial task to
build dark energy model whose EoS can cross the phantom divide. To
this end, a lot of
efforts~\cite{r19,r20,r21,r22,r23,r24,r25,r26,r27,r28,r29,
r30,r31,r32,r33,r34,r35,r36,r67,r68} have been made. To name a few,
quintom model, string theory inspired models, vector field models,
crossing the phantom divide in the braneworld models, scalar-tensor
models~\cite{r68}, etc. However, to our knowledge, many of those
models only provide the possibility that $w_{de}$ can cross $-1$.
They do not answer another question, i.e., {\em why does crossing
the phantom divide occur recently? why are we living in an epoch
$w_{de}<-1$?} This can be regarded as the second cosmological
coincidence problem~\cite{r33,r34}.

Although cosmological observations hint that $w_{de}<-1$, however,
there is a subtle tension between observations and theory. For the
canonical scalar phantom model~\cite{r10}, the universe has an
inevitable fate of big rip~\cite{r11}, and the instability is
inherent~\cite{r12}. For the k-essence model with EoS less than
$-1$, the spatial instabilities inevitably arise too~\cite{r37} (see
also~\cite{r18}). In a more general case, it is argued that there is
a direct connection between instability and the violation of null
energy condition (NEC)~\cite{r38,r39}. Some models can evade this
result, at the price of the lack of isotropy of the background and
the presence of superluminal modes~\cite{r40,r41}. Recently, two
seemingly viable models that violate NEC without instability or
other pathological features have been  proposed in different
ways~\cite{r42,r43}. In particular, in~\cite{r43}, a scalar field
coupled with a vector field is used; and the effective Lagrangian
explicitly depends on the vector filed $A_\mu$, which avoids one of
the assumptions of~\cite{r38,r39} that the effective Lagrangian only
depends on $F_{\mu\nu}\equiv\partial_\mu A_\nu-\partial_\nu A_\mu$
and has no dependence on the vector filed $A_\mu$ itself.

Motivated by the work of~\cite{r43}, it is interesting to study the
case of a scalar field coupled with a vector field. In fact, the
vector field has been used in modern cosmology in many cases, see
e.g.~\cite{r44,r34} and references therein. It is worth noting that
comparing to the ones investigated in e.g.~\cite{r69,r70}, the
vector fields used in~\cite{r44,r34} have fairly different
motivations and forms. In~\cite{r28}, a single dynamical scalar
field is coupled with an {\it a priori} non-dynamical background
vector field with a constant zeroth-component. In that model, the
effective EoS can cross the phantom divide $w=-1$. However, the
appearance of the {\it a priori} non-dynamical vector field has no
clear physical motivation.

In~\cite{r45,r46}, Weyl's old idea of scale
invariance~\cite{r47,r48} was resurrected by Cheng in 1988, almost
60 years later (See~\cite{r49} for an independent rediscovery). The
requirement of local scale invariance leads to the existence of  a
completely new vector field, which we call the ``Cheng-Weyl vector
field'' throughout this paper, in honor of the proposer Cheng and
Weyl (a great mathematician and physicist~\cite{r50}). The
Cheng-Weyl vector field only couples to the scalar field and the
gravitational field. It does not interact with other known matters
in the standard model of particle physics, such as quarks, leptons,
gauge mesons, and so on. In particular, it has no interaction with
photons and electrons. So, it is ``dark'' in this sense. As
mentioned above, the fact that Cheng-Weyl vector field naturally
couples to the scalar field makes it very interesting, especially
when the scalar field is considered as dark energy candidate.
Required by the local scale invariance, the potential term of the
scalar field has to be of $\phi^4$ form, while the coupling form
between the Cheng-Weyl vector field and the scalar field is fixed
also, and the form is different from the ones of~\cite{r28,r43}.
Interestingly, the effective Lagrangian also explicitly depends on
the Cheng-Weyl vector field itself naturally. In Sec.~\ref{sec2}, we
will give a brief review of the work of Cheng~\cite{r45,r46}, in
which the Cheng-Weyl vector field was proposed.

In the present work, the (generalized) Cheng-Weyl vector field
coupled with a scalar field and its cosmological application are
investigated. We regard  a mixture of the scalar field and a
so-called ``cosmic triad'' of three mutually orthogonal Cheng-Weyl
vector fields as the dark energy in the universe. We derive the
effective energy density and pressure of the ``mixed'' dark
energy, and the equations of motion for the scalar field and the
Cheng-Weyl vector field respectively. The cosmological evolution
of this ``mixed'' dark energy is studied. We find that the
effective EoS of dark energy can cross the phantom divide
$w_{de}=-1$ in some cases; the first and second cosmological
coincidence problems can be alleviated at the same time in this
model.

This paper is organized as follows. In Sec.~\ref{sec2}, we will
briefly present the main points of the Cheng-Weyl vector field
proposed in~\cite{r45,r46}. In Sec.~\ref{sec3}, the effective energy
density, pressure, and the equations of motion are obtained. In
Sec.~\ref{sec4}, The cosmological evolution of the ``mixed'' dark
energy is investigated by means of dynamical system~\cite{r51}; the
first and second cosmological coincidence problems are discussed.
Finally, a brief conclusion is given in Sec.~\ref{sec5}.

Throughout this paper, we use the units $\hbar=c=1$ and the
notation $\kappa^2\equiv 8\pi G$, and adopt the metric convention
as $(+,-,-,-)$.

%============================= section 2 ===================================

\section{\label{sec2} The Cheng-Weyl vector field}
Following~\cite{r45,r46}, here we give a brief review of the
so-called Cheng-Weyl vector field. The arguments are based on the
local scale invariance. It is important to distinguish the scale
invariance from the gauge invariance. The scale invariance is the
invariance of the action under the change of the {\em magnitude}
rather than the {\em phase} of the fields. To be definite, let us
consider the distance between two neighboring spacetime points,
$ds^2=g_{\mu\nu}dx^\mu dx^\nu$.  We change the scale of the
distance, for instance, changing the unit of length from meter to
inch. With this change, the distance remains the same, but is
measured in a different unit~\cite{r46}. That is,
 \be{eq1}
g_{\mu\nu}\to\tilde{g}_{\mu\nu}=\Lambda^2 g_{\mu\nu},
 \ee
 where
$\Lambda$ is a constant for the global scale invariance, and is a
function of space and time for the local scale invariance. Then we
have $ds^2\to d\tilde{s}^2\equiv \tilde{g}_{\mu\nu}dx^\mu
dx^\nu=\Lambda^2 ds^2$, and $g^{\mu\nu}\to\tilde{g}^{\mu\nu}
=\Lambda^{-2}g^{\mu\nu}$. And,
$\sqrt{|g|}\to\sqrt{|\tilde{g}|}=\Lambda^4\sqrt{|g|}$, where $g$
is the determinant of the metric $g_{\mu\nu}$. So, the action
$S=\int d^4 x\,\sqrt{|g|}\,{\cal L}$ is invariant under the scale
transformations, provided that the Lagrangian density satisfies
\be{eq2}
 {\cal L}\to\tilde{\cal L}=\Lambda^{-4}{\cal L}.
 \ee
 In this case, the forms of all equations in the theory remain the
same.

Let us first see the case of the global scale invariance, i.e.
$\Lambda$ is a constant. The Lagrangian density of a scalar field
is given by
 \be{eq3}
\frac{1}{2}g^{\mu\nu}\partial_\mu\phi\partial_\nu\phi-\lambda\phi^4,
\ee
 where $\lambda$ is a dimensionless constant. The Lagrangian
density of a gauge meson is
 \be{eq4}
-\frac{1}{4}g^{\mu\rho}g^{\nu\sigma}F_{\mu\nu}F_{\rho\sigma},
 \ee
where $F_{\mu\nu}=\partial_\mu A_\nu-\partial_\nu A_\mu$, or
$F^a_{\mu\nu}=\partial_\mu W^a_\nu-\partial_\nu
W^a_\mu-gf^{abc}W^b_\mu W^c_\nu$ (here $g$ is a coupling constant)
for the Yang-Mills theory. The Lagrangian density for a fermion
$\Psi$ coupled with the electromagnetic field and the
gravitational field is
 \be{eq5}
\bar{\Psi}i\gamma^c\varepsilon^\mu_c\left[\partial_\mu+ieA_\mu
-\frac{1}{2}\sigma_{ab}\varepsilon^{b\nu}\left(\partial_\mu\varepsilon^a_\nu
-\Gamma^\rho_{\mu\nu}\varepsilon^a_\rho\right)\right]\Psi,
 \ee
where $\Gamma^\rho_{\mu\nu}=g^{\rho\sigma}(\partial_\mu
g_{\sigma\nu} +\partial_\nu g_{\sigma\mu}-\partial_\sigma
g_{\mu\nu})/2$,
$\sigma^{ab}=(\gamma^a\gamma^b-\gamma^b\gamma^a)/4$, and
$\varepsilon^a_\mu$ is the tetrad satisfying
$\eta_{ab}\varepsilon^a_\mu\varepsilon^b_\nu=g_{\mu\nu}$. It is
easy to verify that the above Lagrangian densities satisfy
Eq.~(\ref{eq2}) under the scale transformation Eq.~(\ref{eq1}) and
 \be{eq6}
\phi\to\tilde{\phi}=\Lambda^{-1}\phi,~~~A_\mu\to\tilde{A}_\mu=A_\mu,~~~
W^a_\mu\to\tilde{W}^a_\mu=W^a_\mu,~~~
\varepsilon^a_\mu\to\tilde{\varepsilon}^a_\mu=\Lambda\varepsilon^a_\mu,~~~
\Psi\to\tilde{\Psi}=\Lambda^{-3/2}\Psi.
 \ee

Next we consider the case with the theory being scale invariant
locally, i.e. $\Lambda$ is a function of space and time. Similar
to the well-known arguments used to deduce the existence of gauge
fields, one can find that a completely new vector field $\Pi_\mu$,
namely the so-called Cheng-Weyl vector field, is required by the
local scale invariance, while the replacements
 \bea &\partial_\mu
g_{\nu\rho}\to\left(\partial_\mu+2f\Pi_\mu\right)g_{\nu\rho},~~~~~~~
\partial_\mu g^{\nu\rho}\to\left(\partial_\mu-2f\Pi_\mu\right)g^{\nu\rho},\nonumber\\
&\partial_\mu\varepsilon^a_\nu\to\left(\partial_\mu+f\Pi_\mu\right)\varepsilon^a_\nu,~~~~~~~
\partial_\mu\varepsilon_a^\nu\to\left(\partial_\mu-f\Pi_\mu\right)\varepsilon_a^\nu,\nonumber\\
&\partial_\mu\phi\to\left(\partial_\mu-f\Pi_\mu\right)\phi,~~~~~~~
\partial_\mu\Psi\to\left(\partial_\mu-\frac{3}{2}f\Pi_\mu\right)\Psi,\label{eq7}
 \eea
 are also required in these Lagrangian densities, where $f$ is
a dimensionless constant. One can verify that these Lagrangian
densities with the replacements Eq.~(\ref{eq7}) satisfy
Eq.~(\ref{eq2}) under the scale transformation Eqs.~(\ref{eq1}),
(\ref{eq6}) and
 \be{eq8}
\Pi_\mu\to\tilde{\Pi}_\mu=\Pi_\mu-\frac{1}{f}\,\partial_\mu\ln\Lambda.
\ee
 With the replacements Eq.~(\ref{eq7}), the Lagrangian density
of the scalar field Eq.~(\ref{eq3}) becomes
 \be{eq9}
\frac{1}{2}g^{\mu\nu}\left(\partial_\mu-f\Pi_\mu\right)\phi\left(\partial_\nu
-f\Pi_\nu\right)\phi-\lambda\phi^4.
 \ee
  Thus, the scalar field is
coupled with $\Pi_\mu$ naturally by the scale invariance. However,
with the replacements Eq.~(\ref{eq7}), the Lagrangian densities of
the gauge meson and the fermion, i.e. Eqs.~(\ref{eq4}) and
(\ref{eq5}) respectively, need not to be altered, since the terms
involving $\Pi_\mu$ completely cancel one another~\cite{r45,r46}.
Therefore, we conclude that the gauge meson and the fermion do not
couple with $\Pi_\mu$. With identical arguments, the quarks and
leptons etc. do not couple with $\Pi_\mu$  as well. After all, we
would like to mention that the Lagrangian density of $\Pi_\mu$
itself~\cite{r45},
 \be{eq10}
-\frac{1}{4}g^{\mu\rho}g^{\nu\sigma}Y_{\mu\nu}Y_{\rho\sigma},
 \ee
also satisfies Eq.~(\ref{eq2}) under the transformations
Eqs.~(\ref{eq1}), (\ref{eq6}) and (\ref{eq8}), where
 \be{eq11}
Y_{\mu\nu}\equiv\partial_\mu\Pi_\nu-\partial_\nu\Pi_\mu.
 \ee

We close this section with some remarks. First, required by the
local scale invariance, the potential term of the scalar field has
to be of $\phi^4$ form, as in Eqs.~(\ref{eq3}) and~(\ref{eq9}).
Second, it is easy to see from Eq.~(\ref{eq9}) that the coupling
form between the Cheng-Weyl vector field $\Pi_\mu$ and the scalar
field $\phi$ is fixed naturally. Note that the form is quite
different from the ones considered in~\cite{r28,r43}.  For more
details on the Cheng-Weyl vector field,  please see \cite{r45,r46}.
In addition, one may  also refer to \cite{r60,r61} for relevant
papers on the scale invariance.

%============================= section 3 ===================================

\section{\label{sec3} Applying the Cheng-Weyl vector field to cosmology}
In modern cosmology, the scalar field is used extensively. Actually,
the scalar field is one of the leading dark energy candidates. If
the nature respects the local scale invariance, the so-called
Cheng-Weyl vector field must exist, and couples to the scalar field
inherently. If the scalar field is indeed the cause driving the
accelerated expansion of the universe, we argue that the dark energy
should be a mixture of the scalar field and the Cheng-Weyl vector
field, which can be considered as the partner of the scalar field.
This seems quite plausible when the fact that the Cheng-Weyl vector
field does not interact with other known matters (so, it is ``dark''
to them) is taken into account. Therefore it is quite interesting to
study the cosmological consequence of the Cheng-Weyl vector.

We begin with the action
\be{eq12}
S=S_{grav}+S_\cw+S_m,
\ee
where $S_{grav}$ and $S_m$ are the actions for gravitational field and matters respectively,
and
\be{eq13}
S_\cw=\int d^4 x\,\sqrt{-g}\,{\cal L}_\cw.
\ee
Naively, one may write the Lagrangian density ${\cal L}_\cw$ as
$${\cal L}_\cw=\frac{1}{2}g^{\mu\nu}\left(\partial_\mu\phi-f\Pi_\mu\phi\right)
\left(\partial_\nu\phi-f\Pi_\nu\phi\right)-\lambda\phi^4-\frac{1}{4}g^{\mu\alpha}g^{\nu\beta}
Y_{\mu\nu}Y_{\alpha\beta},$$ directly from Eqs.~(\ref{eq9})
and~(\ref{eq10}). In order to be compatible with homogeneity and
isotropy, the $\Pi_\mu$ can be chosen as $\Pi_\mu=(\Pi_0,0,0,0)$,
where $\Pi_0=\Pi_0(t)$ only depends on the cosmic time~$t$. However,
in this case $Y_{\mu\nu}=0$. From this Lagrangian density ${\cal
L}_\cw$, for the case of homogeneous $\phi$, one finds that
$\Pi_0=\phi=0$ which is not dynamical, and the Lagrangian density
${\cal L}_\cw$ is zero actually. Thus, unfortunately, the naive
approach is not viable.

Enlightened by the work of~\cite{r44} (see also~\cite{r34}), we can
 describe the dark energy as a mixture of a scalar field and a
 so-called ``cosmic triad'' (in the terminology of~\cite{r44}) of
 three mutually orthogonal Cheng-Weyl vector fields. In this case,
 the Lagrangian density ${\cal L}_\cw$ is given by
 \bea
{\cal L}_\cw &=& \sum\limits_{a=1}^3\left[\,\frac{\epsilon}{2}g^{\mu\nu}
\left(\partial_\mu\phi-f\Pi^a_\mu\phi\right)\left(\partial_\nu\phi-f\Pi^a_\nu\phi\right)-\lambda\phi^4
-\frac{\eta}{4}g^{\mu\alpha}g^{\nu\beta}Y^a_{\mu\nu}Y^a_{\alpha\beta}\right]\nonumber\\
&=& \sum\limits_{a=1}^3\left[\,\frac{\epsilon}{2}g^{\mu\nu}\partial_\mu\phi\partial_\nu\phi
-\lambda\phi^4-\epsilon fg^{\mu\nu}\Pi^a_\mu\phi\partial_\nu\phi
+\frac{\epsilon}{2}g^{\mu\nu}f^2\Pi^a_\mu\Pi^a_\nu\phi^2
-\frac{\eta}{4}g^{\mu\alpha}g^{\nu\beta}Y^a_{\mu\nu}Y^a_{\alpha\beta}\right]\nonumber\\
&\equiv& \sum\limits_{a=1}^3{\cal L}^{(a)}_\cw,\label{eq14}
 \eea
where
 \be{eq15}
Y^a_{\mu\nu}\equiv\partial_\mu\Pi^a_\nu-\partial_\nu\Pi^a_\mu,
 \ee
$\epsilon$, $\eta$, $\lambda$ and $f$ being dimensionless constants.
Latin indices label the different Cheng-Weyl vector fields
($a,b,\ldots=1,2,3$) and Greek indices label different spacetime
components ($\mu,\nu,\ldots=0,1,2,3$). Actually, the number of
Cheng-Weyl vector fields is dictated by the number of spatial
dimensions and the requirement of isotropy~\cite{r44,r34}. The Latin
indices are raised and lowered with the flat ``metric''
$\delta_{ab}$. It is worth noting that the Lagrangian density ${\cal
L}_\cw$ in Eq.~(\ref{eq14}) satisfies the requirement of the local
scale invariance.  Note that in Eq.~(\ref{eq14}), we have
generalized the original scalar field to include the cases of
quintessence~($\epsilon=+1$) and phantom~($\epsilon=-1$), while the
Lagrangian density for the Cheng-Weyl vector fields has also been
generalized by introducing the constant $\eta$.

Varying the action (\ref{eq13}) with Eq.~(\ref{eq14}), one can get
the energy-momentum tensor of the ``mixed'' dark energy as
 \be{eq16}
^{(\cw)}T_{\mu\nu}=\frac{2}{\sqrt{-g}}\frac{\delta S_\cw}{\delta
g^{\mu\nu}} =\sum\limits_{a=1}^3\left[-g_{\mu\nu}{\cal L}^{(a)}_\cw
+2\frac{\delta {\cal L}^{(a)}_\cw}{\delta g^{\mu\nu}}\right],
 \ee
where
 \be{eq17} \frac{\delta {\cal L}^{(a)}_\cw}{\delta g^{\mu\nu}}
=\epsilon\left(\frac{1}{2}\partial_\mu\phi\partial_\nu\phi-f\Pi^a_\mu\phi\partial_\nu\phi
+\frac{1}{2}f^2 \Pi^a_\mu\Pi^a_\nu\phi^2\right)
-\frac{\eta}{2}g^{\sigma\beta}Y^a_{\mu\sigma}Y^a_{\nu\beta}.
 \ee
From the action~(\ref{eq13}) with Eq.~(\ref{eq14}), one can also
obtain the equations of motion for $\phi$ and $\Pi_\mu^a$, namely
\be{eq18}
\epsilon\,\partial_\mu\left[\sqrt{-g}\sum\limits_{a=1}^3\left(g^{\mu\nu}\partial_\nu\phi
-fg^{\mu\nu}\Pi^a_\nu\phi\right)\right]
=\sqrt{-g}\sum\limits_{a=1}^3\left(-4\lambda\phi^3-\epsilon
fg^{\mu\nu}\Pi^a_\mu\partial_\nu\phi +\epsilon f^2
g^{\mu\nu}\Pi^a_\mu\Pi^a_\nu\phi\right),
 \ee
 and
 \be{eq19}
\eta\,\partial_\mu\left(\sqrt{-g}\,
g^{\mu\alpha}g^{\nu\beta}Y^a_{\alpha\beta}\right)
=\epsilon\sqrt{-g}\left(fg^{\mu\nu}\phi\partial_\mu\phi-g^{\mu\nu}f^2
\Pi^a_\mu\phi^2\right),
 \ee
 respectively. We consider a spatially
flat Friedmann-Robertson-Walker (FRW) universe with metric
 \be{eq20}
ds^2=dt^2-a^2(t)d{\bf x}^2,
 \ee
 where $a(t)$ is the scale factor. In
this work, we assume the scalar field is homogeneous, namely
$\phi=\phi(t)$. Similar to~\cite{r44}, an ansatz for the Cheng-Weyl
vectors is
 \be{eq21}
 \Pi^b_\mu=\delta^b_\mu\,\Pi (t)\cdot a(t).
 \ee
 Thus, the three Cheng-Weyl vectors point in mutually orthogonal spatial
directions, and share the same time-dependent length, i.e.
$\Pi^{a2}\equiv-g^{\mu\nu}\Pi_\mu^a \Pi^a_\nu=\Pi^2(t)$. Hence, the
equations of motion~(\ref{eq18}) and~(\ref{eq19}) become,
respectively,
 \be{eq22}
\epsilon\left(\ddot{\phi}+3H\dot{\phi}\right)+\epsilon f^2
\Pi^2\phi+4\lambda\phi^3=0,
 \ee
 and
 \be{eq23}
\eta\left[\ddot{\Pi}+3H\dot{\Pi}+\left(H^2+\frac{\ddot{a}}{a}\right)\Pi\right]+\epsilon
f^2 \Pi\phi^2=0,
 \ee
 where $H\equiv\dot{a}/a$ is the Hubble
parameter, and a dot denotes the derivative with respect to the
cosmic time~$t$. From Eqs.~(\ref{eq16}), (\ref{eq14})
and~(\ref{eq17}), we find that
 \be{eq24}
 \rho_\cw={}^{(\cw)}T^0_{\ 0}
=\frac{3}{2}\,\epsilon\,\dot{\phi}^2+3\lambda\phi^4
+\frac{3}{2}\,\epsilon f^2
\Pi^2\phi^2+\frac{3}{2}\eta\left(\dot{\Pi}+H\Pi\right)^2,
 \ee
 where
$\rho_\cw$ is the energy density of dark energy, while
 \be{eq25}
^{(\cw)}T^0_{\ i}=0,
 \ee
 and
 \be{eq26}
 ^{(\cw)}T^i_{\ j}=\sum\limits_{a=1}^3 {}^{(a)}T^i_{\ j},
 \ee
 where
 \be{eq27}
^{(a)}T^i_{\ j} =-\left[\frac{\epsilon}{2}\dot{\phi}^2-\lambda\phi^4
-\frac{\epsilon}{2}f^2\Pi^2\phi^2 +\frac{\eta}{2}\left(\dot{\Pi}
+H\Pi\right)^2\right]\delta^i_{\ j}-\left[\epsilon f^2\Pi^2\phi^2
-\eta\left(\dot{\Pi}+H\Pi\right)^2\right]\delta^i_{\ a}\,
\delta^a_{\ j}.
 \ee

It is worth noting here that in fact, even adopting  the ``cosmic
 triad'' of three mutually orthogonal Cheng-Weyl vector fields,
 Lagrangian~(\ref{eq14}) is not invariant under $SO(3)$ rotation
 in the internal space. Thus, the energy-momentum tensor is not
 strictly diagonal. One can see this point by noting that
 $^{(\cw)}T^i_{\ 0}\not=0$, due to the second term in the right hand
 side of Eq.~(\ref{eq17}). To overcome this inconsistence with the
 isotropy, the spatial volume-average procedure has to be employed
 here as done in~\cite{r63,r64}. In those two papers the authors considered
 the nonlinear electromagnetic field as the source driving the
 accelerated expansion of the universe.  There, in order to obtain an
 energy-momentum tensor consistent with the FRW metric, the spatial
 volume-average procedure in {\em the large scale} \cite{r65,r66} has
 been used. By using this procedure, the {\em spatial volume-averaged}
 non-diagonal components of the energy-momentum tensor become zero,
 namely $\left\langle^{(\cw)}T^i_{\ 0}\right\rangle=0$, while the diagonal
 components are kept unchanged. Therefore, in our model, the
 energy-momentum tensor is a {\em spatial volume-averaged} one on
 {\em the cosmological scale}. In this way, the energy-momentum
 tensor is compatible with isotropy.

The corresponding pressure of dark energy is given by
 \be{eq28}
 p_\cw=-\,{}^{(\cw)}T^i_{\ i}
=\frac{3}{2}\,\epsilon\,\dot{\phi}^2-3\lambda\phi^4
-\frac{\epsilon}{2}f^2\Pi^2\phi^2+\frac{\eta}{2}\left(\dot{\Pi}+H\Pi\right)^2.
\ee
 The Friedmann equation and Raychaudhuri equation
read, respectively,
 \be{eq29}
H^2=\frac{\kappa^2}{3}\rho_{tot}=\frac{\kappa^2}{3}\left(\rho_\cw+\rho_m\right),
 \ee
 and
 \be{eq30}
\dot{H}=-\frac{\kappa^2}{2}\left(\rho_{tot}+p_{tot}\right)
=-\frac{\kappa^2}{2}\left(\rho_\cw+\rho_m+p_\cw+p_m\right),
 \ee
where $p_m$ and $\rho_m$ are the pressure and energy density of the
matters, respectively.

From Eqs.~(\ref{eq24}) and~(\ref{eq28}), we obtain
\be{eq31}
\rho_\cw+p_\cw=3\epsilon\,\dot{\phi}^2
+\epsilon f^2\Pi^2\phi^2+2\eta\left(\dot{\Pi}+H\Pi\right)^2.
\ee
Obviously, the EoS of dark energy $w_\cw\equiv p_\cw/\rho_\cw$ is always larger than
$-1$ for the case of $\epsilon>0$ and $\eta>0$, while $w_\cw<-1$ for the case of
$\epsilon<0$ and $\eta<0$. Crossing the phantom divide is impossible for both cases.
However, for the case of $\epsilon$ and $\eta$ having opposite signs, $w_\cw$ can be
larger than or smaller than $-1$. Of course, for this case, crossing the phantom divide is
possible.

In addition, from Eqs.~(\ref{eq24}) and~(\ref{eq28}), we have
\bea
\dot{\rho}_\cw+3H\left(\rho_\cw+p_\cw\right) &=& 3\dot{\phi}\left[\epsilon\left(\ddot{\phi}
+3H\dot{\phi}\right)+\epsilon f^2 \Pi^2\phi+4\lambda\phi^3\right]\nonumber\\
& &+\,3\left(\dot{\Pi}+H\Pi\right)\left\{\eta\left[\ddot{\Pi}+3H\dot{\Pi}
+\left(2H^2+\dot{H}\right)\Pi\right]+\epsilon f^2 \Pi\phi^2\right\}.\label{eq32}
\eea
Noting that $\ddot{a}/a=H^2+\dot{H}$ and the equations of motion for $\phi$
and $\Pi$, i.e. Eqs.~(\ref{eq22}) and~(\ref{eq23}), it is easy to see that the
energy conservation equation of dark energy holds, namely,
$\dot{\rho}_\cw+3H\left(\rho_\cw+p_\cw\right)=0$.

%============================= section 4 ===================================

\section{\label{sec4} Dynamical system and cosmological evolution}
In this section, we investigate the cosmological evolution of the
``mixed'' dark energy by means of dynamical system~\cite{r51}. Our
main aim is to see whether this model can alleviate the coincidence
problems. Like many considerations in the literature, we allow the
existence of interaction between the dark energy and the background
matter (usually the cold dark matter). The cases of the scalar field
and vector field interacting with background matter are studied
extensively, see, for
examples,~\cite{r52,r53,r54,r55,r56,r57,r58,r34}. Although the
Cheng-Weyl vector field does not interact with the {\em known}
matters in the particle physics standard model, nothing precludes
the possibility of the Cheng-Weyl vector field
 interacting with the cold dark matter, since the nature of cold dark matter
 is also {\em unknown} so far.

%============================= section 4.1 ===================================

\subsection{\label{sec4.1} Dynamical system}
We assume that the dark energy and background matter interact through
interaction terms $C$ and $Q$, namely
\bea
&\dot{\rho}_\cw+3H\left(\rho_\cw+p_\cw\right)=-C-Q,\label{eq33}\\
&\dot{\rho}_m+3H\left(\rho_m+p_m\right)=C+Q,\label{eq34}
\eea
which keep the total energy conservation equation
$\dot{\rho}_{tot}+3H\left(\rho_{tot}+p_{tot}\right)=0$. The background matter is described by
a perfect fluid with  barotropic equation of state
\be{eq35}
p_m=w_m\rho_m\equiv (\gamma-1)\rho_m,
\ee
where the barotropic index $\gamma$ is a dimensionless constant and satisfies
$0<\gamma\leq 2$. In particular, $\gamma=1$ and $4/3$ correspond to dust matter and
radiation, respectively. Due to the interaction terms $C$ and $Q$, the equations of motion for
$\phi$ and $\Pi$, namely Eqs.~(\ref{eq22}) and~(\ref{eq23}), are altered. Seeing from
Eqs.~(\ref{eq32}) and~(\ref{eq33}), they are given by
\bea
3\dot{\phi}\left[\epsilon\left(\ddot{\phi}
+3H\dot{\phi}\right)+\epsilon f^2 \Pi^2\phi+4\lambda\phi^3\right]=-C,\nonumber\\
3\left(\dot{\Pi}+H\Pi\right)\left\{\eta\left[\ddot{\Pi}+3H\dot{\Pi}
+\left(2H^2+\dot{H}\right)\Pi\right]+\epsilon f^2
\Pi\phi^2\right\}=-Q,\label{eq36} \eea respectively.
Following~\cite{r51,r52,r53,r54,r55,r56}, we introduce following
dimensionless variables \be{eq37}
x\equiv\frac{\kappa\dot{\phi}}{\sqrt{2}H},~~~~y\equiv\frac{\kappa\dot{\Pi}}{\sqrt{2}H},~~~~
z\equiv\frac{\kappa\sqrt{\rho_m}}{\sqrt{3}H},~~~~u\equiv\frac{\phi}{H},~~~~
v\equiv\frac{\kappa\Pi}{\sqrt{2}}. \ee
With the help of
Eqs.~(\ref{eq29}), (\ref{eq30}), (\ref{eq24}) and~(\ref{eq28}), the
evolution equations~(\ref{eq36}) and~(\ref{eq34}) can be rewritten
as a dynamical system~\cite{r51}, i.e. \bea
&&x^\prime=\left(\Theta_1-3\right)x-2\sqrt{2}\,\lambda\,\epsilon^{-1}u\Theta_2
-\sqrt{2}\,f^2v^2u^3\,\Theta_2^{-1}-C_1,\label{eq38}\\
&&y^\prime=\left(\Theta_1-3\right)y+\left(\Theta_1-\epsilon\,\eta^{-1}f^2u^2-2\right)v
-Q_1,\label{eq39}\\
&&z^\prime=\left(\Theta_1-\frac{3}{2}\gamma\right)z+C_2+Q_2,\label{eq40}\\
&&u^\prime=u\left(\Theta_1+\sqrt{2}\,xu\Theta_2^{-1}\right),\label{eq41}\\
&&v^\prime=y,\label{eq42}
\eea
where
\be{eq43}
C_1\equiv\frac{\kappa^2C}{6\epsilon H^3 x},~~~~
Q_1\equiv\frac{\kappa^2 Q}{6\eta H^3 (y+v)},~~~~
C_2\equiv\frac{z\,C}{2H\rho_m},~~~~Q_2\equiv\frac{z\,Q}{2H\rho_m},
\ee
and
\be{eq44}
\Theta_1\equiv-\frac{\dot{H}}{H^2}=3\epsilon x^2+\epsilon f^2u^2v^2+2\eta (y+v)^2
+\frac{3}{2}\gamma z^2,
\ee
\be{eq45}
\Theta_2\equiv\left[\frac{1-z^2-\epsilon x^2-\epsilon f^2u^2v^2
-\eta (y+v)^2}{\lambda}\right]^{1/2},
\ee
a prime denotes derivative with respect to the so-called $e$-folding time $N\equiv\ln a$.

The fractional energy densities of the background matter and dark
energy are given by
 \be{eq46}
\Omega_m\equiv\frac{\kappa^2\rho_m}{3H^2}=z^2, \ee and \be{eq47}
\Omega_\cw\equiv\frac{\kappa^2\rho_\cw}{3H^2}=\epsilon
x^2+\lambda\kappa^2 H^2u^4 +\epsilon f^2u^2v^2+\eta (y+v)^2,
 \ee
respectively. On the other hand, from Eq.~(\ref{eq29}), one has
\be{eq48} \Omega_\cw=1-z^2. \ee Hence, from Eqs.~(\ref{eq47})
and~(\ref{eq48}), one can find out
 \be{eq49} \kappa
H=\frac{\Theta_2}{u^2},
 \ee
 where $\Theta_2$ is given byEq.~(\ref{eq45}). The effective EoS
 of the whole system is
 \be{eq50}
 w_{eff}\equiv\frac{p_{tot}}{\rho_{tot}}
 =\Omega_\cw w_\cw+\Omega_m w_m,
 \ee
 where $w_\cw\equiv p_\cw/\rho_\cw$ and $w_m\equiv p_m/\rho_m=\gamma-1$
 are the EoS of dark energy and background matter, respectively.

%============================= section 4.2 ===================================

\subsection{\label{sec4.2} Interaction terms and critical points}
In this subsection, we obtain all critical points of the dynamical
system~(\ref{eq38})--(\ref{eq42}). A critical point $(\bar{x},
\bar{y}, \bar{z}, \bar{u}, \bar{v})$ satisfies the conditions
$\bar{x}^\prime=\bar{y}^\prime=\bar{z}^\prime=\bar{u}^\prime=\bar{v}^\prime=0$.
Before giving the particular interaction terms $C$ and $Q$, let us
first find the general features of the critical points of dynamical
system~(\ref{eq38})--(\ref{eq42}). From Eq.~(\ref{eq42}) and
$\bar{v}^\prime=0$, it is easy to see that \be{eq51} \bar{y}=0. \ee
If this dynamical system has some critical points, their
corresponding $\bar{x}$, $\bar{y}$, $\bar{z}$, $\bar{u}$, and
$\bar{v}$ should be constants. Therefore, from Eqs.~(\ref{eq49})
and~(\ref{eq45}), the corresponding Hubble parameter
$H=\bar{H}=const$. From Eq.~(\ref{eq44}), \be{eq52} \bar{\Theta}_1=0
\ee follows. Substituting into Eq.~(\ref{eq41}), $\bar{u}^\prime=0$
requires \be{eq53} \bar{x}=0, \ee since
$\bar{u}^2\bar{\Theta}_2^{-1}=(\kappa\bar{H})^{-1}\not=0$. Hence,
from
 Eqs.~(\ref{eq52}), (\ref{eq44}), (\ref{eq51}) and~(\ref{eq53}),
 we have
 \be{eq54}
 \left(\epsilon f^2\bar{u}^2+2\eta\right)\bar{v}^2+\frac{3}{2}\gamma\bar{z}^2=0.
 \ee
 So, $\epsilon f^2\bar{u}^2+2\eta<0$ is required for non-vanishing real $\bar{z}$ and $\bar{v}$.
 By using Eqs.~(\ref{eq38})--(\ref{eq40}) and Eqs.~(\ref{eq51})--(\ref{eq54}),
 $\bar{x}^\prime=\bar{y}^\prime=\bar{z}^\prime=0$ become, respectively,
 \bea
 \sqrt{2}\,\bar{u}\left(\epsilon\,\bar{\Theta}_2\right)^{-1}\left(2-\frac{\gamma}{2}\bar{z}^2\right)
 +\bar{C}_1&=&0,\label{eq55}\\
 \left(\epsilon f^2\bar{u}^2+2\eta\right)\bar{v}+\eta\,\bar{\Theta}_1&=&0,\label{eq56}\\
 \bar{C}_2+\bar{Q}_2-\frac{3}{2}\gamma\bar{z}&=&0,\label{eq57}
 \eea
where \be{eq58}
\bar{\Theta}_2=\left[\frac{1+\left(\frac{3}{2}\gamma-1\right)\bar{z}^2+\eta\bar{v}^2}{\lambda}\right]^{1/2},
\ee which comes from Eqs.~(\ref{eq45}), (\ref{eq51}), (\ref{eq53}),
and~(\ref{eq54}). Then, one can find out the remaining $\bar{z}$,
$\bar{u}$ and $\bar{v}$ from Eqs.~(\ref{eq54})--(\ref{eq57}).
Obviously, only three of them are independent of each other.

So far, the above results are independent of particular interaction
terms $C$ and $Q$. To find out $\bar{z}$, $\bar{u}$ and~$\bar{v}$,
we have to choose proper $C$ and $Q$ here. The interaction forms
extensively considered in the literature
(see~\cite{r51,r52,r53,r54,r55,r56,r57,r58,r34,r32} for instance)
are
\begin{eqnarray*}
C\propto H\rho_m,\ H\rho_{tot},\ H\rho_\cw,\ \kappa\rho_m\dot{\phi},\ \ldots\\
Q\propto H\rho_m,\ H\rho_{tot},\ H\rho_\cw,\ \kappa\rho_m\dot{\Pi},\ \ldots
\end{eqnarray*}
Noting that Eqs.~(\ref{eq53}), (\ref{eq37}) and the definition of $C_1$ in Eq.~(\ref{eq43}), we
have to choose
\be{eq59}
C=\alpha\kappa\rho_m\dot{\phi},
\ee
to avoid the divergence of $\bar{C}_1$ in Eq.~(\ref{eq55}), where $\alpha$ is a dimensionless
constant. In this case, from Eq.~(\ref{eq43}), one has
\be{eq60}
C_1=\frac{\alpha z^2}{\sqrt{2}\,\epsilon},~~~~~~~C_2=\frac{\alpha xz}{\sqrt{2}}.
\ee
From Eqs.~(\ref{eq53}) and~(\ref{eq60}), we find that $\bar{C}_2=0$ in Eq.~(\ref{eq57}). Noting
that Eqs.~(\ref{eq51}), (\ref{eq53}), (\ref{eq46}) and the definition of $Q_2$ in Eq.~(\ref{eq43}), we
cannot choose $Q\propto H\rho_m$ or $\kappa\rho_m\dot{\Pi}$, to avoid the solution $\bar{z}=0$
from Eq.~(\ref{eq57}), since our main aim is to alleviate the coincidence problems. Therefore, we
choose {\bf Case~(I)}~$Q=3\beta H\rho_\cw$ or {\bf Case~(II)}~$Q=3\sigma H\rho_{tot}$, where
$\beta$ and $\sigma$ are dimensionless constants.

%============================= section 4.2.1 ===================================

\subsubsection*{\label{sec4.2.1} {\bf Case~(I)}~$Q=3\beta H\rho_\cw$}
In this case, from Eq.~(\ref{eq43}), one has
 \be{eq61}
Q_1=\frac{3\beta (1-z^2)}{2\eta (y+v)},~~~~~~~
Q_2=\frac{3}{2}\beta\left(z^{-1}-z\right).
 \ee
 Noting that
$\bar{C}_2=0$ in Eq.~(\ref{eq57}), we find out
 \be{eq62}
\bar{z}=\sqrt{\frac{\beta}{\beta+\gamma}}.
 \ee
 One can check that
Eq.~(\ref{eq56}) is equivalent to Eq.~(\ref{eq57}) for this case.
Then, one can find out $\bar{u}$ and $\bar{v}$ from
Eqs.~(\ref{eq54}) and~(\ref{eq55}), by using Eqs.~(\ref{eq51}),
(\ref{eq53}) and~(\ref{eq62}). We do not present them here, since
the final results are involved and tedious. One can work them out
with the help of Mathematica. Instead we would like to give several
particular examples to support our statement. {\bf Example
(I.1)},~for parameters $\gamma=1$, $\epsilon=1$, $\eta=-1$,
$\alpha=5$, $\beta=3/5$, $\lambda=0.1$ and $f=5$, we have
$\bar{u}=-0.246341$, $\bar{v}=\pm 1.07927$. {\bf Example (I.2)},~for
parameters $\gamma=1$, $\epsilon=1$, $\eta=-1$, $\alpha=5$,
$\beta=1/2$, $\lambda=0.1$ and $f=5$, we find that
$\bar{u}=-0.249803$, $\bar{v}=\pm 1.06605$.

From Eqs.~(\ref{eq46}), (\ref{eq48}), and~(\ref{eq62}), the
fractional energy densities of background matter and dark energy
are given by
 \be{eq63}
\Omega_m=\frac{\beta}{\beta+\gamma},~~~~~~~\Omega_\cw=\frac{\gamma}{\beta+\gamma},
 \ee
 respectively. For reasonable $\Omega_m$ and $\Omega_\cw$, it
is easy to see that $\beta>0$ is required. As mentioned above, at
the critical point $(\bar{x}, \bar{y}, \bar{z}, \bar{u},
\bar{v})$, the Hubble parameter $H=\bar{H}=const$. From
Eq.~(\ref{eq30}), this means
 \be{eq64}
 w_{eff}=-1.
 \ee
 From Eq.~(\ref{eq50}), we find that the EoS of the dark energy is
 given by
 \be{eq65}
 w_\cw=-1-\beta.
 \ee
 Obviously, $w_\cw<-1$.

%============================= section 4.2.2 ===================================

\subsubsection*{\label{sec4.2.2} {\bf Case~(II)}~$Q=3\sigma H\rho_{tot}$}
In this case, the corresponding $Q_1$ and $Q_2$ read \be{eq66}
Q_1=\frac{3\sigma}{2\eta (y+v)},~~~~~~~Q_2=\frac{3}{2}\sigma z^{-1},
\ee respectively. Solving Eq.~(\ref{eq57}) with $\bar{C}_2=0$, we
get \be{eq67} \bar{z}=\sqrt{\frac{\sigma}{\gamma}}. \ee Again, one
can check that Eq.~(\ref{eq56}) is equivalent to Eq.~(\ref{eq57})
for this case. Then, one can find out $\bar{u}$ and $\bar{v}$ from
Eqs.~(\ref{eq54}) and~(\ref{eq55}), by using Eqs.~(\ref{eq51}),
(\ref{eq53}) and~(\ref{eq67}). Once again, we do not present the
long and involved expressions here. We only give some particular
examples. {\bf Example~(II.1)},~for parameters $\gamma=1$,
$\epsilon=1$, $\eta=-1$, $\alpha=4$, $\sigma=1/3$, $\lambda=0.1$ and
$f=3$, we find that $\bar{u}=-0.410797$, $\bar{v}=\pm 1.01934$. {\bf
Example~(II.2)},~for parameters $\gamma=1$, $\epsilon=1$, $\eta=-1$,
$\alpha=5$, $\sigma=0.3$, $\lambda=0.1$ and $f=7$, we have
$\bar{u}=-0.180804$, $\bar{v}=\pm 1.06307$.

From Eqs.~(\ref{eq46}), (\ref{eq48}), and~(\ref{eq67}), the
fractional energy densities of background matter and dark energy are
given by, respectively, \be{eq68}
\Omega_m=\frac{\sigma}{\gamma},~~~~~~~\Omega_\cw=1-\frac{\sigma}{\gamma},
\ee which requires $0<\sigma<\gamma$. Following a similar argument,
we have $w_{eff}=-1$ also. And then, from Eq.~(\ref{eq50}), we find
that the EoS of the dark energy is given by \be{eq69}
w_\cw=-1-\frac{\sigma\gamma}{\gamma-\sigma}, \ee which is also
smaller than $-1$.

%============================= section 4.3 ===================================

\subsection{\label{sec4.3} Stability analysis}
In this subsection, we discuss the stability of these critical
points. An attractor is one of the stable critical points of the
autonomous system. To study the stability of these critical points,
we substitute linear perturbations $x\to\bar{x}+\delta x$,
$y\to\bar{y}+\delta y$, $z\to\bar{z}+\delta z$, $u\to\bar{u}+\delta
u$ and $v\to\bar{v}+\delta v$ about the critical point $(\bar{x},
\bar{y}, \bar{z}, \bar{u}, \bar{v})$ into the dynamical system
Eqs.~(\ref{eq38})--(\ref{eq42}) and linearize them. We get the
evolution equations for the fluctuations as \bea \delta x^\prime &=&
\left(\bar{\Theta}_1-3\right)\delta x+\bar{x}\delta\Theta_1
-2\sqrt{2}\lambda\epsilon^{-1}\left(\bar{u}\delta\Theta_2+\bar{\Theta}_2\delta
u\right)
+\sqrt{2} f^2\bar{v}^2\bar{u}^3\bar{\Theta}_2^{-2}\delta\Theta_2\nonumber\\
& &-\sqrt{2}f^2\left(3\bar{u}^2\bar{v}^2\delta u+2\bar{u}^3\bar{v}\delta v\right)\bar{\Theta}_2^{-1}
-\delta C_1,\label{eq70}\\
\delta y^\prime &=& \left(\bar{\Theta}_1-3\right)\delta y+\bar{y}\delta\Theta_1
+\left(\bar{\Theta}_1-\epsilon\eta^{-1}f^2\bar{u}^2-2\right)\delta v
+\left(\delta\Theta_1-2\epsilon\eta^{-1}f^2\bar{u}\delta u\right)\bar{v}-\delta Q_1,\label{eq71}\\
\delta z^\prime &=& \left(\bar{\Theta}_1-\frac{3}{2}\gamma\right)\delta z
+\bar{z}\delta\Theta_1+\delta C_2+\delta Q_2,\label{eq72}\\
\delta u^\prime &=& \bar{u}\left[\delta\Theta_1-\sqrt{2}\bar{x}\bar{u}\bar{\Theta}_2^{-2}\delta\Theta_2
+\sqrt{2}\left(\bar{x}\delta u+\bar{u}\delta x\right)\bar{\Theta}_2^{-1}\right]
+\left(\bar{\Theta}_1+\sqrt{2}\bar{x}\bar{u}\bar{\Theta}_2^{-1}\right)\delta u,\label{eq73}\\
\delta v^\prime &=& \delta y,\label{eq74}
\eea
where $\delta\Theta_1$, $\delta\Theta_2$, $\delta C_1$, $\delta C_2$, $\delta Q_1$
and~$\delta Q_2$ are the linear perturbations coming from $\Theta_1$, $\Theta_2$, $C_1$,
$C_2$, $Q_1$ and~$Q_2$, respectively. The five eigenvalues of the coefficient matrix of
the above equations determine the stability of the corresponding critical point.

Now, we work out $\delta\Theta_1$, $\delta\Theta_2$, $\delta C_1$,
$\delta C_2$, $\delta Q_1$ and~$\delta Q_2$ one by one. From
Eq.~(\ref{eq44}), we get
 \be{eq75}
\delta\Theta_1=6\epsilon\bar{x}\delta x+2\epsilon
f^2\left(\bar{u}^2\bar{v}\delta v +\bar{u}\bar{v}^2\delta
u\right)+4\eta\left(\bar{y}+\bar{v}\right)\left(\delta y+\delta
v\right) +3\gamma\bar{z}\delta z.
 \ee
 From Eq.~(\ref{eq45}), we have
 \be{eq76}
\delta\Theta_2=\left(\lambda\bar{\Theta}_2\right)^{-1}\left[\bar{z}\delta
z+\epsilon\bar{x}\delta x +\epsilon
f^2\left(\bar{u}^2\bar{v}\delta v+\bar{u}\bar{v}^2\delta u\right)
+\eta\left(\bar{y}+\bar{v}\right)\left(\delta y+\delta
v\right)\right].
 \ee
 From Eq.~(\ref{eq60}), it is easy to find that
 \be{eq77}
 \delta C_1=\frac{\sqrt{2}\alpha}{\epsilon}\bar{z}\delta z,~~~~~~~
 \delta C_2=\frac{\alpha}{\sqrt{2}}\left(\bar{x}\delta z
 +\bar{z}\delta x\right).
 \ee
Then, we obtain $\delta Q_1$ and $\delta Q_2$ for
{\bf Case~(I)} and {\bf Case~(II)} respectively, since they depend
on the particular form of $Q$. For the {\bf Case~(I)}, from
Eq.~(\ref{eq61}), we obtain
 \be{eq78}
 \delta Q_1=-\frac{3\beta\left(1-\bar{z}^2\right)}{2\eta\left(\bar{y}
 +\bar{v}\right)^2}\left(\delta y
 +\delta v\right)-\frac{3\beta\bar{z}}{\eta\left(\bar{y}
 +\bar{v}\right)}\delta z,~~~~~~~
 \delta Q_2=-\frac{3}{2}\beta\left(\bar{z}^{-2}+1\right)\delta z.
 \ee
 For the {\bf Case~(II)}, from Eq.~(\ref{eq66}), it is easy to get
 \be{eq79}
 \delta Q_1=-\frac{3\sigma}{2\eta}\left(\bar{y}
 +\bar{v}\right)^{-2}\left(\delta y+\delta v\right), ~~~~~~~
 \delta Q_2=-\frac{3}{2}\sigma\bar{z}^{-2}\delta z.
 \ee

We substitute the critical point $(\bar{x}, \bar{y}, \bar{z},
\bar{u}, \bar{v})$ and $\bar{\Theta}_1=0$ as well as
$\bar{\Theta}_2$ given by Eq.~(\ref{eq58}) into
Eqs.~(\ref{eq70})--(\ref{eq74}) with
Eqs.~(\ref{eq75})--(\ref{eq79}). The five eigenvalues of the
coefficient matrix of these equations determine the stability of the
corresponding critical point. We find that for both {\bf
Case~(I)}~$Q=3\beta H\rho_\cw$ and {\bf Case~(II)}~$Q=3\sigma
H\rho_{tot}$, the critical points can exist and are stable in proper
parameter spaces, respectively. In other words, {\em they are late
time attractors}. Needless to say, the particular parameter spaces
for the existence and stability of these critical points are
considerably involved and tedious. Since our main aim here is just
to point out the fact that it can exist and is stable, we do not
present those very involved expressions for the corresponding
parameter space. Of course, one can work out with the help of
Mathematica. Here, we just give several particular examples to
support our statement. For the {\bf Example~(I.1)} of the {\bf
Case~(I)} mentioned above, the corresponding eigenvalues are
$\{-3.10805+i\, 3.37335,\ -3.10805-i\, 3.37335,\ -3.43776,\
-1.49442,\ -0.134622\}$; for the {\bf Example~(I.2)}, the
corresponding eigenvalues are $\{-3.10108+i\, 3.12107,\ -3.10108-i\,
3.12107,\ -2.95079,\ -1.67829,\ -0.108711\}$; for the {\bf
Example~(II.1)} of the {\bf Case~(II)} mentioned above, the
corresponding eigenvalues are $\{-2.68308+i\, 2.6312,\ -2.68308-i\,
2.6312,\ -2.42269,\ -1.6597,\ -0.032674\}$; for the {\bf
Example~(II.2)}, the corresponding eigenvalues are $\{-2.68735+i\,
3.12848,\ -2.68735-i\, 3.12848,\ -1.97246+i\, 0.723607,\ -1.97246
-i\, 0.723607,\ -0.0785649\}$. Obviously, they are all stable.

%============================= section 4.4 ===================================

\subsection{\label{sec4.4} The first and second cosmological coincidence problems}
Here, following the argument in Sec.~VI of~\cite{r34}, we briefly
show that the first and second cosmological coincidence problems can
be alleviated at the same time in our model. As is well known, the
approach most frequently used to alleviate the cosmological
coincidence problem is the scaling attractor(s) in the dynamical
system (see~\cite{r51,r52,r53, r54,r55,r56,r57,r58,r34,r32} for
examples). The most desirable feature of dynamical system is that
the whole system will eventually evolve to its attractors, having
nothing to do with the initial conditions. Therefore, fine-tuning is
unnecessary.

As is explicitly shown in this work, {\em all} stable attractors
have the desirable properties, namely, their corresponding
$\Omega_m$ and $\Omega_\cw$ are comparable, while $w_\cw<-1$. As
mentioned in the end of Sec.~\ref{sec3} of the present paper, for
the case of $\epsilon$ and $\eta$ have opposite signs, $w_\cw$ can
be larger than or smaller than $-1$. Thus, crossing the phantom
divide is possible. For a fairly wide range of initial conditions
with $w_\cw>-1$, the universe will eventually evolve to the scaling
attractor(s) with $w_\cw<-1$, while the corresponding $\Omega_m$ and
$\Omega_\cw$ are comparable. Thus, it is not strange that we are
living in an epoch when the densities of dark energy and matter are
comparable, and the EoS of dark energy is smaller than $-1$. In this
sense, the first and second cosmological coincidence problems are
alleviated at the same time in our model.

%============================= section 5 ===================================

\section{\label{sec5} Conclusion and discussions}

In summary,  the (generalized) Cheng-Weyl vector field coupled
with a scalar field and its cosmological application are
investigated in the present work. In our model, the dark energy is
described as a mixture of a scalar field and a so-called ``cosmic
triad'' of three mutually orthogonal Cheng-Weyl vector fields. We
derive the effective energy density and pressure of the ``mixed''
dark energy, and the equations of motion for the scalar field and
the Cheng-Weyl vector field, respectively, by using the spatial
volume-average procedure. The cosmological evolution of this
``mixed'' dark energy is studied. We find that the effective EoS
can cross the phantom divide $w_{de}=-1$ in the case of $\epsilon$
and $\eta$ having opposite signs. The first and second
cosmological coincidence problems can be alleviated at the same
time in our model. On the other hand, it is easy to see that {\em
all} stable attractors have $w_{eff}=-1$. Although the EoS of dark
energy $w_\cw$ can be smaller than $-1$, the big rip never appears
in this model. The fate of our universe is an inflationary phase,
in which the Hubble parameter $H=\bar{H}=const$.

we finish this paper with some remarks. In this work, to cross the
phantom divide, $\epsilon$ and $\eta$ should have opposite signs.
This means that either the scalar field or the vector field has a
negative sign of its kinetic term. As is argued in~\cite{r71}, these
ghost-type fields are possible and viable. Second, one might notice
that from Eqs.~(\ref{eq63}), (\ref{eq65}), (\ref{eq68})
and~(\ref{eq69}), at the attractors, the appropriate value
$\Omega_m\sim 0.25$ at $\beta\sim 1/3$ or $\sigma\sim 1/4$ results
$w_\cw\sim -4/3$, for $\gamma=1$. However, this is not inconsistent
with the observations~\cite{r72}. The situation becomes more
comfortable when one is aware of the possibility that the universe
perhaps nears but has not reached the attractors. Finally, at the
attractors, $\Omega_m$ is constant and not equal to zero, thanks to
the interaction between the vector fields and the background matter.
In fact, the interaction between dark energy and dark matter can be
constrained by the cosmological observations, see~\cite{r73,r67} for
examples. However, there is still a long way to obtain some strict
constraints on this interaction.

%============================= acknowledgments ===================================

\section*{ACKNOWLEDGMENTS}
We thank the anonymous referee for useful comments and suggestions,
which help us to improve this paper. We are grateful to
Prof.~Hung~Cheng for his impressive talk on~\cite{r45,r46} given at
ITP, July 2004, which inspired the present work. We also thank
Prof.~Xin-Min~Zhang for drawing our attention to the work of
Rubakov~\cite{r43}, and thank Prof.~Yue-Liang~Wu for comments on
this paper. H.W. is grateful to Zong-Kuan~Guo, Yun-Song~Piao,
Bo~Feng, Yan~Chai, Ding-Fang~Zeng, Xun~Su, Wei-Shui~Xu, Hui~Li,
Li-Ming~Cao, Da-Wei~Pang, Xin~Zhang, Qing-Guo~Huang, Yi~Zhang,
Qi~Guo, Hong-Sheng~Zhang, Hong-Bao~Zhang, Hua~Bai, Jia-Rui~Sun,
Ding~Ma, Ya-Wen~Sun, Xin-Qiang~Li and Hao~Ma for kind help and
useful discussions. This work was supported in part by China
Postdoctoral Science Foundation, and a grant from Chinese Academy of
Sciences (No.~KJCX3-SYW-N2), and grants from NSFC, China
(No.~10325525, No.~90403029 and No.~10525060).

\renewcommand{\baselinestretch}{1.1}

%============================= references ===================================

\end{document}